# Data Hiding in Video using Triangularization LSB Technique

**Subhashri Acharya[1], Pramita Srimany[2], Sanchari Kundu[3], JayatiGhosh Dastidar[4]**
[1]St. Xavier's College, Kolkata, India,subhashri.acharya@gmail.com
[2]St. Xavier's College, Kolkata, India, pramitasrimany@gmail.com
[3]St. Xavier's College, Kolkata, India,sancharikundu15@gmail.com
[4]St. Xavier's College, Kolkata, India, j.ghoshdastidar@sxccal.edu

**ABSTRACT**

The importance of data hiding in the field of Information Technology is a widely accepted. The challenge is to be able to pass information in a manner that the very existence of the message is unknown in order to repel attention of the potential attacker. Steganography is a technique that has been widely used to achieve this objective. However Steganography is often found to be lacking when it comes to hiding bulk data. Attempting to hide data in a video overcomes this problem because of the large sized cover object (video) as compared to an image in the case of steganography. This paper attempts to propose a scheme using which data can be hidden in a video. We focus on the Triangularization method and make use of the Least Significant Bit (LSB) technique in hiding messages in a video.

## 1. INTRODUCTION

**Information security**, sometimes shortened to **InfoSec**, is the practice of defending information from unauthorized access, use, disclosure, disruption, modification, inspection, recording or destruction. It is a general term that can be used regardless the form the data may take (e.g. electronic, physical).
Information security must protect information throughout the life span of the information, from the initial creation of the information on through to the final disposal of the information. To fully protect the information during its lifetime, each component of the information processing system must have its own protection mechanisms. The building up, layering on and overlapping of security measures is called defense in depth.

Broadly we can classify the data security in two parts:
1. Cryptography
2. Steganography

In cryptography, the structure of a message is scrambled to make it meaningless and unintelligible unless the decryption key is available. It makes no attempt to disguise or hide the encoded message. Basically, cryptography offers the ability of transmitting information between persons in a way that prevents a third party from reading it. Cryptography is the practice and study of techniques for secure communication in the presence of third parties. Data modification is done here with some algorithms. A 'key' is also used to make data more secure. Without knowing that 'key', the third party will not be able to break the security and therefore will not get the 'secured' data. Cryptography is basically a text based technique.

In contrast, steganography is the technique for data hiding [3]. It is the art and science of writing hidden messages in such a way that no one, apart from the sender and intended recipient, suspects the existence of the message, a form of security through obscurity. The word *steganography* is of Greek origin and means "concealed writing" from the Greek words *steganos* meaning "covered or protected", and *graphei*meaning "writing".

The advantage of cryptography is that it hides the message and privacy is safe. Along with this, no one would be able to know what it says unless there's a key to the code. One can write whatever he/she wants and however we want (i.e., any theme any symbol for the code) to keep our code a secret.
The advantage of steganography is that messages do not attract attention to themselves. Steganography protects both messages and communicating parties. It includes the concealment of information within computer files. In digital steganography, electronic communications may include steganographic coding inside of a transport layer, such as a document file, image file, program or protocol. Media files are ideal for steganographic transmission because of their large size [2]. The idea of hiding messages in digital images is a long practiced approach.

We describe a method which uses a key to protect the message hidden in a **video** instead of an image.We use the goodness of both the worlds of cryptography and steganography. In addition, the use of video makes the process of extracting the hidden message almost impossible to achieve unless the hiding process is known.

## 2. DATA HIDING USING STEGANOGRAPHY

### 2.1 Overview

The basic model of steganography consists of Carrier, Message and password. Carrier is also known as cover-object, in which the message is embedded and serves to hide the presence of the message.

Basically, the model for steganography is shown in Figure 1:

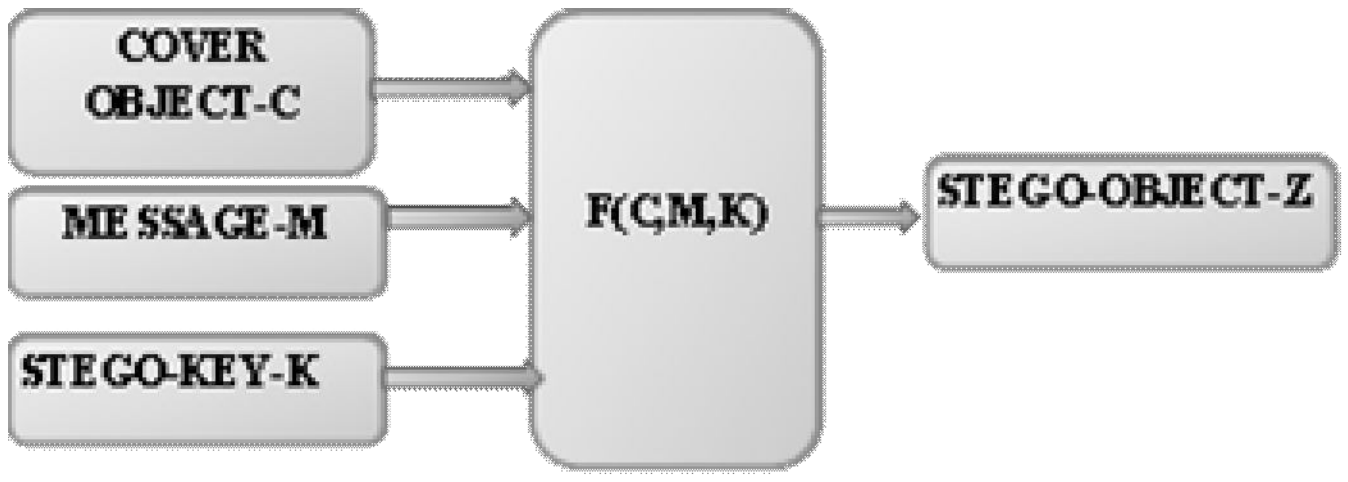

**Figure 1:** Basic Steganography Model

Message, M is the data that the sender wishes to keep confidential. It can be plain text, cipher text, another image, or anything that can be embedded in a bit stream such as a copyright mark, a covert communication, or a serial number. Password is known as the stego-key, K which ensures that only recipient who knows the corresponding decoding key will be able to extract the message from a cover-object. The cover-object, C with the secretly embedded message is then called the Stego-object, Z [2]. The function, F(C, M, K) is the process used for hiding the message.

Recovering the message from a stego-object requires the cover-object itself and a corresponding decoding key if a stego-key was used during the encoding process. The original image may or may not be required in most applications to extract the message [6].

Several suitable carriers may be used as the cover-object [8]:

- Network packets/segments such as IP, TCP and UDP.
- Audio files using digital audio formats such as wav, midi, avi, mpeg, mpi and voc.
- Images file such as bmp, gif and jpg, where they can be both color and gray-scale.
- Video files using digital video formats such as avi, mp4.
- File and Disk that can hide and append files by using the slack space.

### 2.2 Techniques of Steganography

Over the past few years, numerous steganography techniques that embed hidden messages in multimedia objects have been proposed. Common approaches include:

- Least significant bit insertion (LSB).
- Masking and filtering.
- Transform techniques.

Least significant bits (LSB) insertion is a simple approach to embedding information in an image file. The simplest steganographic techniques embed the bits of the message directly into least significant bit plane of the cover-image in a deterministic sequence. Modulating the least significant bit does not result in human-perceptible difference because the amplitude of the change is small [5].

Masking and filtering techniques are usually restricted to 24 bits and gray scale images. In this data is hidden by marking an image, in a manner similar to paper watermarks.

Transform techniques embed the message by modulating coefficients in a transform domain, such as the Discrete Cosine Transform (DCT), Discrete Fourier Transform (DFT), or Wavelet Transform. These methods hide messages in significant areas of the cover-image, which make them more robust to attacks.

We propose the use of the LSB method along with triangularization[1] in this paper. Triangularization is used to modify the message to be hidden and the LSB method is used for hiding that modified message into a video file[7].

## 3. DETAILED DESIGN

### 3.1 Steganography Algorithm

This algorithm is used to hide a message in a video. The algorithm works by first encrypting the message to be hidden. The encryption process is a Triangularizationmethod (described later). This encrypted message is then hidden in a cover object, in this case a video. In order to do this frames are extracted from the video and the encrypted (triangularized) image is hidden in them using the LSB substitution technique. The algorithm has been described below:

*Input: Original Video, Message*
*Output: Stego-video*
*Method:*

1. Take the input message.
2. The input message is masked using masking.
3. To encrypt the message, triangularization method is performed.
4. Read the original video and create the demo video as empty.
5. A key is generated.
6. Frames are extracted from the video.
7. According to the key, frames are selected for hiding. The arrangements of frames are random.
8. Hide the encrypted message into the LSB's of original video frame-wise.
9. Thus, we get a cipher video (stego video).

### 3.2 Desteganography Algorithm

This algorithm is used to extract the hidden message from the stego video. This process is the reverse of the steganography algorithm.

*Input:stego video, key*
*Output: Original message*
*Method:*

1. Decrypt the message file using the cipher video and the key.
2. Decrypt the message using detriangularization method.
3. Bit XOR is performed to unmask the original message.
4. Hence, we get the original message input by the user.

### 3.3 Triangularization/De-Triangularization Algorithm

Initially an n bit data string is taken. After this, we consider the steps discussed below.

1. The data string initialized is taken as it is.
2. Bit wise XOR operation is performed of all the bits; the iteration process is continued until the data string is reduced to a single bit.
3. The LSB's from the data string obtained from each of the iterations and are then joined together and taken as the new output.
4. If we take the new output as the data string, and perform the above mentioned steps, i.e, step1-step4, we get the original output.

The implementation of the above algorithm is shown using a data string example (Table 1).
The input string is 10001110.

**Table 1: Illustration of Triangularization method**

| | | | | | | | | | | | | | | | |
|---|---|---|---|---|---|---|---|---|---|---|---|---|---|---|---|
| 1 | | 0 | | 0 | | 0 | | 1 | | 1 | | 1 | | **0** | ↙ |
| | 1 | | 0 | | 0 | | 1 | | 0 | | 0 | | **1** | ↙ | |
| | | 1 | | 0 | | 1 | | 1 | | 0 | | **1** | ↙ | | |
| | | | 1 | | 1 | | 0 | | 1 | | **1** | ↙ | | | |
| | | | | 0 | | 1 | | 1 | | **0** | ↙ | | | | |
| | | | | | 1 | | 0 | | **1** | ↙ | | | | | |
| | | | | | | 1 | | **1** | ↙ | | | | | | |
| | | | | | | | **0** | ↙ | | | | | | | |

The output string is 01110110. The first row of the above table is the 8 bit input string 10001110. We compute the XOR of adjacent bits and generate the second row. This process is iteratively continued till we reach the last row (7$^{th}$. row). For n bits n-1 iterations will be required. The least significant bit is extracted at every step and joined, starting the first row to generate the output string. The same has been highlighted in red in Table 1 (as indicated by the arrows).

The novelty of this scheme lies in the fact that this method is a reversible operation. In order to generate the original bit string the Triangularization method is once again applied on the output string as demonstrated as follows (Table 2):

**Table 2: Illustration of Reverse Triangularization method**

| | | | | | | | | | | | | | | | |
|---|---|---|---|---|---|---|---|---|---|---|---|---|---|---|---|
| ↖ | **0** | | 1 | | 1 | | 1 | | 0 | | 1 | | 1 | | 0 |
| | ↖ | **1** | | 0 | | 0 | | 1 | | 1 | | 0 | | 1 | |
| | | ↖ | **1** | | 0 | | 1 | | 0 | | 1 | | 1 | | |
| | | | ↖ | **1** | | 1 | | 1 | | 1 | | 0 | | | |
| | | | | ↖ | **0** | | 0 | | 0 | | 1 | | | | |
| | | | | | ↖ | **0** | | 0 | | 1 | | | | | |
| | | | | | | ↖ | **0** | | 1 | | | | | | |
| | | | | | | | ↖ | **1** | | | | | | | |

However, during the De-Triangularization process the original bit string is generated by extracting the most significant bits starting the last row (last iteration) in reverse (indicated by arrows) and joined to generate the original data string 10001110 [4].

### 4. DISCUSSION

The scheme proposed by us is simple and easy to implement. We now proceed to discuss the advantages and disadvantages of this scheme.

#### 4.1 Advantages

1. **Lossless Process:** The method is lossless as demonstrated above. The process can successfully retrieve every data bit that is hidden in the video.

2. **Random key numbers:** One to one mapping is used to generate the key randomly. Each character is converted into a distinct number after encryption and it is not supposed to get same number for a particular character in one run. Key numbers are distinct and non-overlapping in the same run as well as different run. So intruders will not be able to predict the message even if they manage to get the key.

3. **Many-to-many mapping makes stronger:** Encryption technique with masking has the property of many-to-many mapping from plain text domain to cipher text which make "relative frequency of English letter" - approach impossible to break the cipher.

4. **Not Format specified:** All video formats are acceptable. Video format specification is not a bound.

#### 4.2 Disadvantages

1. **Lossless Process:** Encrypted video file is too big to be a cipher even for a small message.

2. **No sound is there in stego video:** The hiding process does not take into account the audio component of the video.

3. **Possibility of same character occurrence**: It is expected that generated random numbers in key will be distinct. No two numbers in a key will be same at a particular run. But it is not guaranteed for a large message, and any two random numbers may be similar.

### 5. CONCLUSION

In this paper we propose a method to hide data in a secure way. We use both cryptography and steganography technique so that it becomes difficult for the attacker to get the message or data. We use masking and triangularization techniques to encrypt data which makes the data more secure. Using masking, the mapping becomes many-to-many so it is difficult for the attacker to guess the actual data. After hiding the plain video and cipher both are looks similar as we hide data in LSB position so that no one can guess which one is the cipher video. As we hide data in lossless video so that file size remains same or reduced without losing the quality of the video. The method can be extended or modified to adopt other known techniques of steganography.